\documentclass[epj]{webofc}
\usepackage[utf8]{inputenc}
\usepackage[varg]{txfonts}   % Web of Conferences font
\usepackage{booktabs}
\usepackage{xcolor}
\definecolor{darkred}{rgb}{0.4,0.0,0.0}
\definecolor{darkgreen}{rgb}{0.0,0.4,0.0}
\definecolor{darkblue}{rgb}{0.0,0.0,0.4}
\usepackage[bookmarks,linktocpage,colorlinks,
    linkcolor = darkred,
    urlcolor  = darkblue,
    citecolor = darkgreen]{hyperref}
\usepackage{subfigure}
\usepackage{graphicx}
\usepackage{amsmath}

\usepackage{amssymb}
\usepackage{rotating}
\usepackage[nodisplayskipstretch]{setspace}
\everydisplay\expandafter{\the\everydisplay\setstretch{1}}
\usepackage{color,hangcaption,hhline,psfrag,rotating,amssymb}
\wocname{EPJ Web of Conferences}
\woctitle{Lattice2017}
\begin{document}
%%%%%%%%%%%%%%%%%%%%%%%%%%%%%%%%%%%%%%%%%%%%%%%%%%%%%%%%%%%%%%%%%%%%%%%%%%%%%
%
\selectlanguage{english}
%----------------------------------------------------------------------------
\title{%
$D \rightarrow Kl\nu$ semileptonic decay using lattice QCD with HISQ at physical pion masses
}
%----------------------------------------------------------------------------
\author{%
\firstname{Bipasha} \lastname{Chakraborty}\inst{1,2}\fnsep\thanks{Acknowledges financial support by U.S. Department of Energy under grants No. DE-AC05-06OR23177}\fnsep\thanks{Speaker, \email{bipasha@jlab.org} } \and
\firstname{Christine} \lastname{Davies}\inst{2} \and
\firstname{Jonna}  \lastname{Koponen}\inst{3}\and 
\firstname{G Peter}  \lastname{Lepage}\inst{4}%\fnsep\thanks{Speaker, \email{bipasha@jlab.org} }
% etc.
}
%----------------------------------------------------------------------------
\institute{%
Jefferson Lab, 12000 Jefferson Avenue, Newport News, Virginia 23606, USA
\and
SUPA, School of Physics and Astronomy, University of Glasgow, Glasgow, G12 8QQ, UK
\and
INFN, Sezione di Roma Tor Vergata, Via della Ricerca Scientifica 1, 00133 Roma RM, Italy
\and
Laboratory for Elementary-Particle Physics, Cornell University, Ithaca, New York 14853, USA
}
%----------------------------------------------------------------------------
\abstract{%
 % Insert your English abstract here, preferably the original one from indico.
  %If this is a joint proceeding containing several contributions, please
  %indicate all speakers in the footnotes. These proceedings are in one-column
  %format.
  
The quark flavor sector of the Standard Model is a fertile ground to look for new physics effects through a unitarity test of the Cabbibo-Kobayashi-Maskawa (CKM) matrix. We present a lattice QCD calculation of the scalar and the vector form factors (over a large $q^2$ region including $q^2=0$) associated with the $D \rightarrow Kl\nu$ semi-leptonic decay. This calculation will then allow us to determine the central CKM matrix element, $V_{cs}$ in the Standard Model, by comparing the lattice QCD results for the form factors and the experimental decay rate. This form factor calculation has been performed on the $N_f=2+1+1$ MILC HISQ ensembles with the physical light quark masses.
}
%----------------------------------------------------------------------------
\maketitle
%----------------------------------------------------------------------------
\section{Introduction}\label{sec:intro}

The flavour changing weak interactions between quarks via emission of $W$ bosons can be parametrised in terms of the Cabbibo-Kobayashi-Maskawa (CKM) unitary matrix in the Standard Model given by~\cite{Cabibbo:1963yz,Kobayashi:1973fv}
\begin{equation}
V_{CKM} = \begin{bmatrix} V_{ud}&V_{us}&V_{ub}\\V_{cd}&V_{cs}&V_{cb}\\V_{td}&V_{ts}&V_{tb} \end{bmatrix}.
\end{equation}
%This matrix also represents the strength of the weak interaction. 
Precise and independent determination of each of the CKM matrix elements is crucial to test the Standard Model stringently and any deviation from unitarity would signal the existence of physics beyond the Standard Model.

The uncertainties in the unitarity checks of the second row and second column of the CKM matrix are dominated by that of $|V_{cs}|$, the central CKM matrix element. This element is calculated from the studies of the leptonic and semileptonic meson decays involving charged flavour changing current from $c$ to $s$ by combining the experimental decay rate with the vector form factor calculated from lattice QCD~\cite{Cabibbo:1963yz,Kobayashi:1973fv}. The best experimental result to date is achieved by combining the experimental data from BaBar~\cite{Aubert:2007wg}, Belle~\cite{Widhalm:2006wz}, BES~\cite{Liu:2012bn} and CLEO~\cite{PhysRevD.80.032005}. However, in the present scenario, the uncertainty in $|V_{cs}|$ is dominated by the lattice uncertainty in the form factors. %For precise determination of $|V_{cs}|$ we need to achieve lattice result of the form factors with the similar precision. %Simultaneously it is also possible to test the internal structure of the mesons in details by studying these decay processes.      

Here, we present a calculation of the semileptonic $D\rightarrow K l \nu$ decay on the $N_f = 2+1+1$ lattices generated by MILC using highly improved staggered quark (HISQ) formalism~\cite{Eduardo}, which is an improvement over HPQCD's previous work reported in~\cite{Na:2010uf,Koponen:2013tua}. In contrast to the work done by Fermilab lattice and MILC collaboration in the reference~\cite{Primer:2015qpz}, we have calculated both the scalar and vector form factors of this decay over the whole range of kaon momentum instead of only at the maximum kaon momentum. A similar study~\cite{Lubicz:2017syv} has been recently done using twisted mass fermions. %However, I have only used the scalar form factor at the zero momentum exchange (from $D$ to $W$ boson) limit to extract $|V_{cs}|$ for the purpose of this dissertation, though 
%Currently we in future this study will be extended to include all experimental momentum squared bins and their correlations, following~\cite{Koponen:2013tua}.            
% Do not forget to mention truncated solver for charm and naik epsilon table
\begin{figure}[tbh]
\caption{The diagram represents the three-point correlator for the $D\rightarrow Kl\nu$ semileptonic decay (on the top) and the two-point correlators for the $D$ and $K$ mesons (at the bottom).}
\label{fig:tasteint}
\centering
\includegraphics[width=0.35\textwidth]{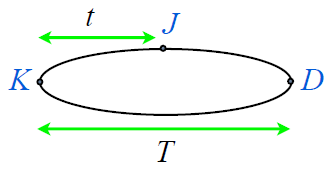}\\
\includegraphics[width=0.35\textwidth]{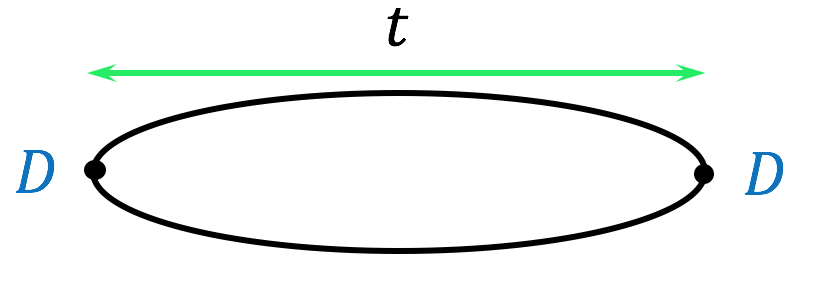}
\includegraphics[width=0.35\textwidth]{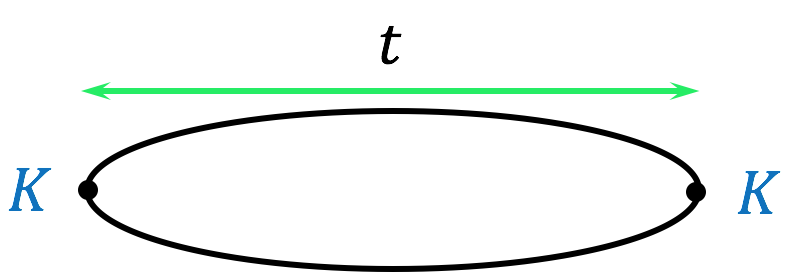}
%\includegraphics[width=0.45\textwidth]{./diagrams/DK2.png} 
%\caption{The diagram represents the three-point correlator for the $D\rightarrow Kl\nu$ semileptonic decay (on the top) and the two-point correlators for the $D$ and $K$ mesons (at the bottom).}
%\label{fig:tasteint}
\end{figure}

%--------------------------------------------------------------

\section{Formalism}

The matrix element of the $D \rightarrow Kl\nu$ semileptonic decay via the charged electroweak current gets contribution only from the vector current. The vector current matrix element can be parametrised in terms of the scalar and vector form factors $f_0(q^2)$ and $f_+(q^2)$, and can be written as - 
%\begin{center}
\begin{eqnarray}
Z_{V,t}\times <K^-|V^\mu|D^0> &=& f_+^{D \rightarrow K(q^2)} [p_D^\mu + p_K^\mu - \frac{M^2_D - M^2_K}{q^2}q^\mu] \nonumber\\
                & & + f_0^{D \rightarrow K(q^2)} \frac{M^2_D - M^2_K}{q^2}q^\mu  
                \label{eq:D2Kff} 
\end{eqnarray} 
%\end{center}
where, $q^\mu = p^\mu_D - p^\mu_K$ is the exchanged 4-momentum from $D$ with a 4-momentum $p^\mu_D$ to $K$ with a 4-momentum $p^\mu_K$ and carried away by the $W$ boson. $M_D$ amd $M_K$ are the masses of the $D$ and $K$ mesons respectively. In our setup, we give momentum to the strange quark inside the $K$ meson. 

The local vector current using the HISQ formalism is not conserved, and therefore requires a renormalisation factor $Z_V$ to obtain the continuum result. We determine this renormalisation by using the partially conserved vector current (PCVC) relation~\cite{Na:2010uf} and also the scalar current amplitude \\$<K^-|S|D^0> = <K^-|\bar{\psi_s}\psi_c|D^0>$. %In the continuum scalar and vector currents obey the PCVC relation written as

The scalar current amplitude is parametrised as
\begin{equation}
<K^-|S|D^0> = \frac{M^2_D - M^2_K}{m_{0c} - m{0s}}f_0^{D \rightarrow K}(q^2)\label{eq:scff}
\end{equation}   

Inverting this equation the scalar form factor for different $q^2$ can be extracted. The scalar current using HISQ is absolutely renormalised when multiplied by lattice quark mass~\cite{Na:2010uf}. 

For easier calculation of the $Z_V$ factor and a better signal we have used only the local temporal vector current. For calculating $Z_{V,t}$ we consider the kinematics at $q^2=q^2_{max}={(M_D-M_K)}^2$ i.e. ${\vec{p}}_k=\vec{0}$. Plugging in the $f_0^{D \rightarrow K}(q^2)$ values from equation \ref{eq:scff} into equation \ref{eq:D2Kff} we can extract $Z_{V,t}$. For different kinematic combinations we can extract the vector form factor $f_+^{D \rightarrow K}(q^2)$ by plugging in the $Z_{V,t}$ value and $f_0^{D \rightarrow K}(q^2)$ values from equation \ref{eq:scff} into equation \ref{eq:D2Kff}. In this way we can cover the full physical range of $q^2$ starting from $q^2_{max}$ where the momentum exchange is maximum i.e. $K$ meson is rest to $q^2 = 0$ where $K$ gets the maximum possible momentum in the opposite direction to leptons.

The differential decay rate is dominated by the vector channel in the vanishing lepton mass limit~\cite{Gupta:1993qp} and we get
\begin{equation}
d\Gamma(q^2) = \frac{G^2_F|V_{cs}|^2}{192\pi^3m^3_D}dq^2\lambda(q^2)^{3/2}|f_+(q^2)|^2\label{eq:diffdecayrate}
\end{equation}

Staggered quarks have four tastes running in the correlator loop. To get a non-zero expectation value of the scalar and vector current operator matrix elements, we need to choose correct combinations of operators at the source, sink and current insertion point such that the correlator becomes taste-singlet. For generating the scalar current amplitude, the current carries spin-taste $1\otimes 1$. We keep the spin-taste at the $K$-meson annihilation point the same for both scalar and vector currents as we want to use the same strange propagators in both cases. This end has the spin-taste content $\gamma_5 \otimes \gamma_5$. To nullify the tastes, the simplest choice of operators for the $D$ meson end is also the Goldstone pseudoscalar operator $\gamma_5 \otimes \gamma_5$. 

For the local temporal vector current, at the current insertion point we have used the $\gamma_t \otimes \gamma_t$ operator. As mentioned before, the same $K$ meson propagators are used in this case as well, therefore we have a spin-taste operator $\gamma_5 \otimes \gamma_5$ at this end. Now, to cancel the overall taste the simplest operator choice at the $D$ meson end would be the local non-Goldstone operator $\gamma_5\gamma_t \otimes \gamma_5\gamma_t$ which generates a $D$ meson with slightly different mass.  
%This couples to the $J^{PC} = 0^{-+}$ states, therefore, can be used for $D$ meson operators. 
To use these three-point correlators we need to make Goldstone and local non-Goldstone two-point $D$ correlators and Goldstone two-point $K$ correlators as well.

\section{Lattice setup}

We have used publicly available MILC HISQ $N_f=2+1+1$ configurations with three different lattice spacing $\sim 0.09 fm$ (fine), $\sim 0.12 fm$ (coarse), $\sim 0.15 fm$ (very coarse) and the physical values of all of the sea quark masses. The details of these configurations are given in Table \ref{tab:D2Kconf}. 

The values of the time sources $t_0$ have been chosen randomly to reduce autocorrelation and for each configuration multiple values of $t_0$, uniformly placed on the lattice, have been used to get better statistics. To increase the statistics further, multiple values of the source-sink separation $T$ have been used for each $t_0$ value.
\begin{table}                                                                                                                       
\caption{Sets of MILC configurations used with their $\beta = 10/g^2$~\cite{MILCensemble}, $w_0/a$ for $w_0 = 0.1715(9)$ fm fixed from $f_\pi$~\cite{DowdallVus}, $L_s/a$, $L_t/a$, number of configurations $N_{conf}$, number of independent time sources for each configuration $t_0$, multiple values of the source-sink separation $T$ for each $t_0$, (HISQ) sea quark masses- $m_l$ (at physical pion mass), $m_s$ and $m_c$ in lattice units~\cite{MILCensemble}.}    
\centering
%\scalebox{0.85}{
\begin{tabular}{ccccccccccc}
\hline
\hline
Set & $\beta$ & $w_0/a$ & $L_s/a$ & $L_t/a$ & $N_{conf}$ & $t_0$  & $T$  &   $am_{l}^{sea}$ & $am_{s}^{sea}$ & $am_c^{sea}$ \\
\hline
3    & 5.8    & 1.1367(5)  &  36   &  48    &  997   &     16    & 9, 12, 15, 18 &     0.00235        &  0.0647        &  0.831 \\
\hline
8    & 6.0    & 1.4149(6)   &  48  &  64    &  998       &  16   & 12, 15, 18, 21  &   0.00184       &   0.0507        &  0.628\\
\hline
11   & 6.0    & 1.9518(7)   & 64   &  96    &  660       &   8   & 16, 19, 22, 25  &  0.00120       &  0.0363         &  0.432\\
\hline
\hline
\end{tabular}
\label{tab:D2Kconf}
\end{table}
%The adjustable parameters in this calculation are the bare valence quark masses and the $\epsilon$ parameter for HISQ~\cite{Eduardo} to improve the ``naik'' term present in the AsqTad~\cite{Naik:1986bn,PhysRevD.59.074502}. 
The valence light quark mass $am_{l}^{val}$ is taken to be the same as the sea light quark mass $am_{l}^{sea}$ whereas the valence strange quark mass $am_{s}^{val,tuned}$ is tuned~\cite{Chakraborty:2014aca} to give the mass of the $\eta_s$ meson to be $0.6885(22)$ GeV~\cite{Chakraborty:2014aca}. We also tuned the valence charm quark mass $am_c^{val,tuned}$ to get the mass of the $\eta_c$ meson $m_{\eta_c} = 2.9863(27)$ GeV~\cite{Chakraborty:2014aca}. %$\epsilon$ vanishes rapidly for decreasing masses of the quarks, therefore is taken as zero for the light and strange quarks. For heavier charm quarks new improved $\epsilon$ given by

The three-point correlation functions on the lattice have been generated using the ``sequential technique'' shown in Figure \ref{fig:tasteint}. In this set up, zero momentum D meson is created at time $t_0 +T$ on lattice, after it propagates to time $t$ on lattice, the current (scalar or local temporal vector) is inserted at time $t$ which changes the flavor $c$ inside the $D$ meson to flavor $s$ to create a $K$ meson and emits a $W$ boson. %Therefore, 
%In our simulation $s$-quark propagator runs from $t_0$ to $t$ whereas the light quark (spectator quark) propagator runs from $t_0$ to $t_0 + T$. $K$ propagates from time $t$ to time $t_0$ on lattice and is annihilated at that time, where, $t_0 < t < t_0 + T$. The $c$-quark propagator is generated from $t_0 + T$ considering the value of the light quark propagator at this particular time slice as the source of the $c$ propagator. This technique is known as ``sequential technique'' and the propagator generated in this way is called as the ``extended propagator''. 
% diagram sequential techniques
%The inversion for the strange propagators for each of the $\vec{p}_K$s is the most expensive part of this calculation as the used zero momentum light propagators were previously generated by Rachel Dowdall for a separate project. 
%And charm propagators are different for only different $T$s. 
%While generating charm propagators the stopping condition (relative error instead of errors in CG method) has been tested using ``truncated solver'' method to achieve the convergence.

%%%%%%%%%%%%%%%%%%%%%%%%%%%%%%%%%%%%%

\section{Fits and data analysis}

The two point heavy-light $D$ and $K$ meson correlators have the following fit form -  
\begin{equation}
G^{2pt}(t;\vec{p}) = \sum_n a^2_n (e^{-E_nt}+e^{-E_n(T-t)}) + (-1)^t \sum_{no} a^2_{no} (e^{-E_{no}t} + e^{-E_{no}(T-t)}).\label{eq:simple2ptPBCosc}
\end{equation}

Here, both mesons in staggered quark formalism have oscillation in the correlators; $E_n$ represents the energy of the $n-$th excited state whereas $E_{no}$ represents the energy of the $n$-th oscillating state. Similarly, $a_n$ and $a_{no}$ respectively give the non-oscillating and oscillating pieces of the amplitude for the $n$-th state of the meson. We have taken for simplicity $t_0 = 0$ by always shifting the source time in the correlators to the origin of the lattice. In the two-point correlators apart from ground state, other excited states are also present, but we are only interested in the mass, energy and amplitude of the ground state for this calculation. 

The ground state probability of the $D/K$ meson is extracted as
\begin{equation}
{(a_o^D)}^2 = \frac{{|\langle 0|\chi_D|D\rangle|}^2}{2M_D a^3},\hspace{1cm}
{(a_o^K)}^2 = \frac{{|\langle 0|\chi_K|K\rangle|}^2}{2M_K a^3}.\label{eq:2ptampD2K}
\end{equation}

Here, $\chi_D$ and $\chi_K$ are the interpolating operators for the $D$ and the $K$ mesons respectively; $a$ is the lattice spacing.

The three-point correlators (for both the scalar and vector currents) have oscillations at both ends and can be written as - %in equation \ref{eq:simple3ptPBCosc}. % as 
\begin{eqnarray}
G^{3pt}(t;T) & = & \sum_{n_1,n_2} a^K_{n_1} a^D_{n_2} V^{nn}_{n_1n_2}(e^{-E^K_{n_1}t}+e^{-E^D_{n_2}(T-t)})\nonumber\\
             &   & (-1)^t\sum_{n_1o,n_2} a^K_{n_1o} a^D_{n_2} V^{on}_{n_1on_2}(e^{-E^K_{n_1o}t}+e^{-E^D_{n_2}(T-t)})\nonumber\\ 
            &   & (-1)^T\sum_{n_1,n_2o} a^K_{n_1} a^D_{n_2o} V^{no}_{n_1n_2o}(e^{-E^K_{n_1}t}+e^{-E^D_{n_2o}(T-t)})\nonumber\\
            &   & (+1)^{t+T}\sum_{n_1o,n_2o} a^K_{n_1o} a^D_{n_2o} V^{oo}_{n_1on_2o}(e^{-E^K_{n_1o}t}+e^{-E^D_{n_2o}(T-t)}).\label{eq:simple3ptPBCosc}
\end{eqnarray}
Here, following a similar notation, ``nn'', ``no'', ``on'' and ``oo'' represent the non-oscillating/non-oscillating, non-oscillating/oscillating, oscillating/non-oscillating, and oscillating/oscillating states respectively.

%Here, we have taken $0\leq t \leq T$ and $T \ll L_t$ such that any contributions from the meson propagating in the opposite direction from the boundary due to the periodic boundary condition can be neglected. Otherwise, it will tremendously complicate the fit form of the three-point correlator. 

We use multi-exponential Bayesian fitting methods~\cite{Lepagenotes} to simultaneously fit the two-point and three-point correlators for multiple $T$s with all correlations among errors taken into account to extract the three-point amplitude $V^{nn}$. %Since using all $T$ values does not increase the number of fit parameters, but increases the amount of data, we get much better statistics from simultaneous fitting and much more precise ground state observables for the mesons. We use both even and odd $T$ values which further improves the fit uncertainties. 

The ground state nonoscillating-nonoscillating amplitude of the three-point function for any current $J$ is
\begin{equation}
V^{nn}_{00} = \frac{<0|\chi_K|K><K|J|D><D|\chi_D|0>}{(2E_K a^3)a_0^K a_0^D}.\label{eq:3ptampD2K}
\end{equation}
%From equations \ref{eq:3ptampD2K}, \ref{eq:simple3ptPBCosc} and \ref{eq:2ptampD2K} we extract 

%Here, the scalar current amplitude is
%\begin{eqnarray}
%\langle K|S|D\rangle = 2\sqrt{M_DE_K}V_{00}^{\mathrm{nn,scalar}}.\label{eq:scfinal1}
%\end{eqnarray}

%And the vector current amplitude is given by
%\begin{eqnarray}
%\langle K|V|D\rangle = Z_V \times 2\sqrt{M_DE_K}V_{00}^{\mathrm{nn,vector}}.\label{eq:scfinal}
%\end{eqnarray}

%And using equations \ref{eq:scfinal1} and \ref{eq:scff} we extract $f_0(q^2)$. %Plugging in the value of $f_0(q^2)$ in the PCVC relation given by equation \ref{eq:pcvc} we extract the renormalisation factor for the local vector current $Z_V$. 
%Putting the values of $f_0(q^2)$ and $Z_V$ in equation \ref{eq:D2Kff0} we can extract $f_+(q^2)$. 
%For getting stable ground state fit results I have tried up to $n_{\mathrm{exp}}=7$, but in most cases I obtained a stable fit and a good $\chi^2$ from the 3rd exponential.% Following priors and their widths (energy in the units of $GeV$) have been used for fits.

For the first sum in equation \ref{eq:simple2ptPBCosc} (and for $\vec{p}_K = \vec{0}$) we have used the priors as (energy in the units of GeV) 
\begin{eqnarray}
\log(E^K_{(0)}) = \log (0.48(5)),\hspace{0.5cm} \log(E^K_{(n)}-E^K_{(n-1)}) = \log (0.40(20)) \quad (n>0)\nonumber\\
\log(E^D_{(0)}) = \log (1.80(18)),\hspace{0.5cm} \log(E^D_{(n)}-E^D_{(n-1)}) = \log (0.40(20)) \quad (n>0)\nonumber\\
a^K_{(0)} = 0.01(1.0),\hspace{0.5cm} a^D_{(0)} = 0.01(1.0).
\end{eqnarray}

%The notation is the same as used in the previous chapters. 
We have assigned analogous priors for the second sum as well, but with
\begin{align}
\log(Eo^{D/K}_{(0)}) = \log(E^D/K_{(0)}+(0.23,0.12)).
\end{align}

For the three point amplitudes we assign
\begin{align}
V^{nn}_{\mathrm{scalar}} = 0.01(5.0), V^{on}_{\mathrm{scalar}} = 0.01(15.0)\nonumber\\
V^{no}_{\mathrm{scalar}} = 0.01(15.0), V^{oo}_{\mathrm{scalar}} = 0.01(15.0)\nonumber\\
V^{nn}_{\mathrm{vector}} =0.01(10.0), V^{on}_{\mathrm{vector}} =0.01(10.0)\nonumber\\
V^{no}_{\mathrm{vector}} =0.01(10.0), V^{oo}_{\mathrm{vector}} = 0.01(10.0).
\end{align}

The energy priors for other kaon momenta $\vec{p}_K$ are given following the dispersion relation $E^2={\vec{p}_K}^{ 2}+m^2$.

The time range we have used for fitting two-point correlators is $[t_{\mathrm{min}},L_t-t_{\mathrm{min}}]$, where $t_{\mathrm{min}}$ for very coarse, coarse and fine lattices are $3,4$ and $5$ respectively. To fit the three point correlators on very coarse, coarse and fine lattices we have used time ranges $[3,T-3]$, $[4,T-4]$ and $[6,T-6]$ respectively. The fits are generally consistent within a range of $t_{\mathrm{min}}$ values.
%%%%%%%%%%%%%%%%%%%%%%%%%%%%%

\section{Results}

% If possible later: 3pt to 2pt plot, Twisted B C plot

%In this calculation for each ensemble I have first run the $q^2_{\mathrm{max}}$ jobs while I have also generated the zero momentum $D$ and $K$ two-point correlators. I have extracted the meson properties like masses, decay constants separately from only two-point fits and from simultaneous fits of two-point and $q^2_{\mathrm{max}}$ three-point correlators. As discussed before, the simultaneous fits give much better precision in the ground state meson properties and the final results are taken from these fits. 

\begin{figure}
\caption{$M_D$ vs. $n_{\mathrm{exp}}$ plot (left), and $a^D_0$ vs. $n_{\mathrm{exp}}$ plot (right) on the physical coarse lattice.}
\label{fig:nexpM}
\centering
\includegraphics[width=0.48\textwidth]{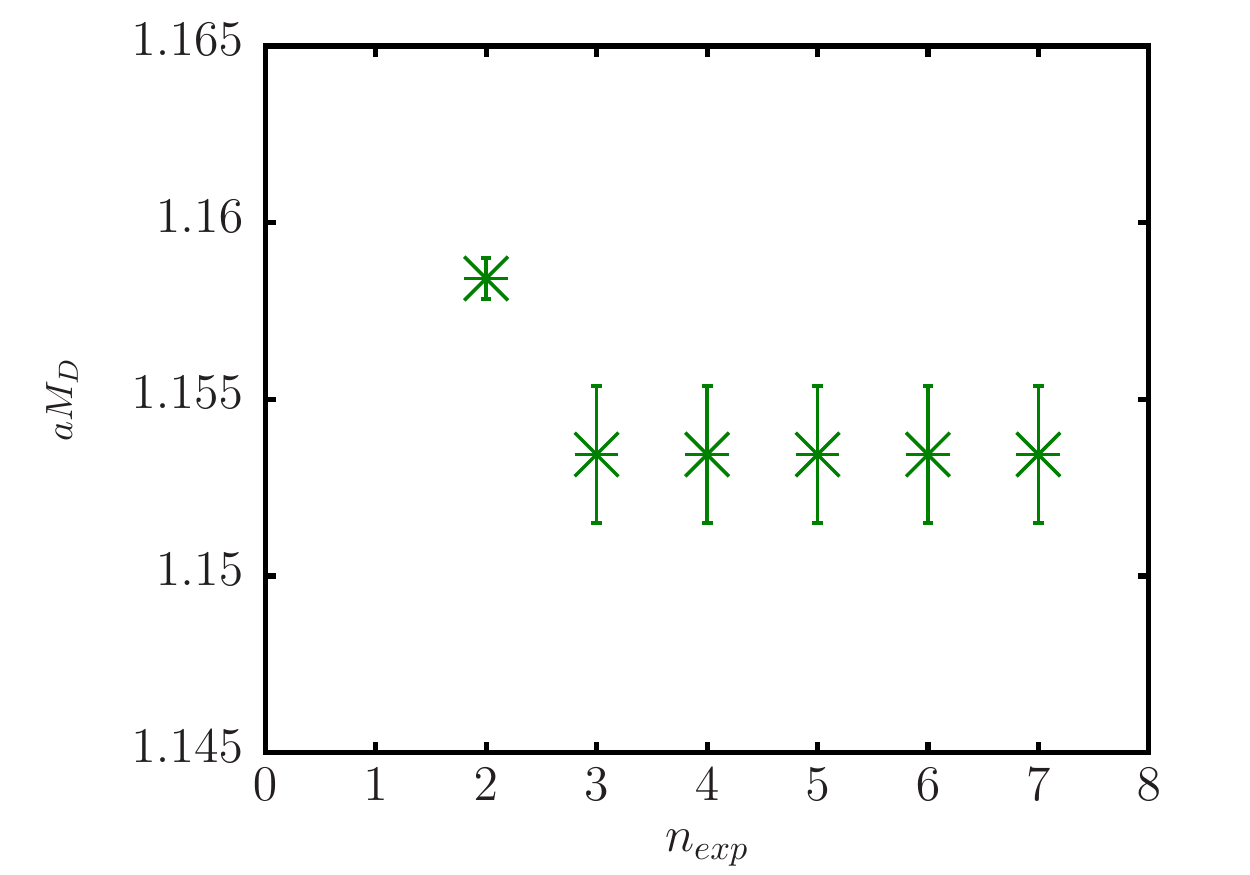}
%\includegraphics[width=0.45\textwidth]{./diagrams/latticediagram.png}                                                                       
%\end{figure}
%\begin{figure}
%\caption{$a^D_0$ vs. $n_{\mathrm{exp}}$ plot on the physical coarse lattice.}
%\label{fig:nexpa0}
%\centering
\includegraphics[width=0.48\textwidth]{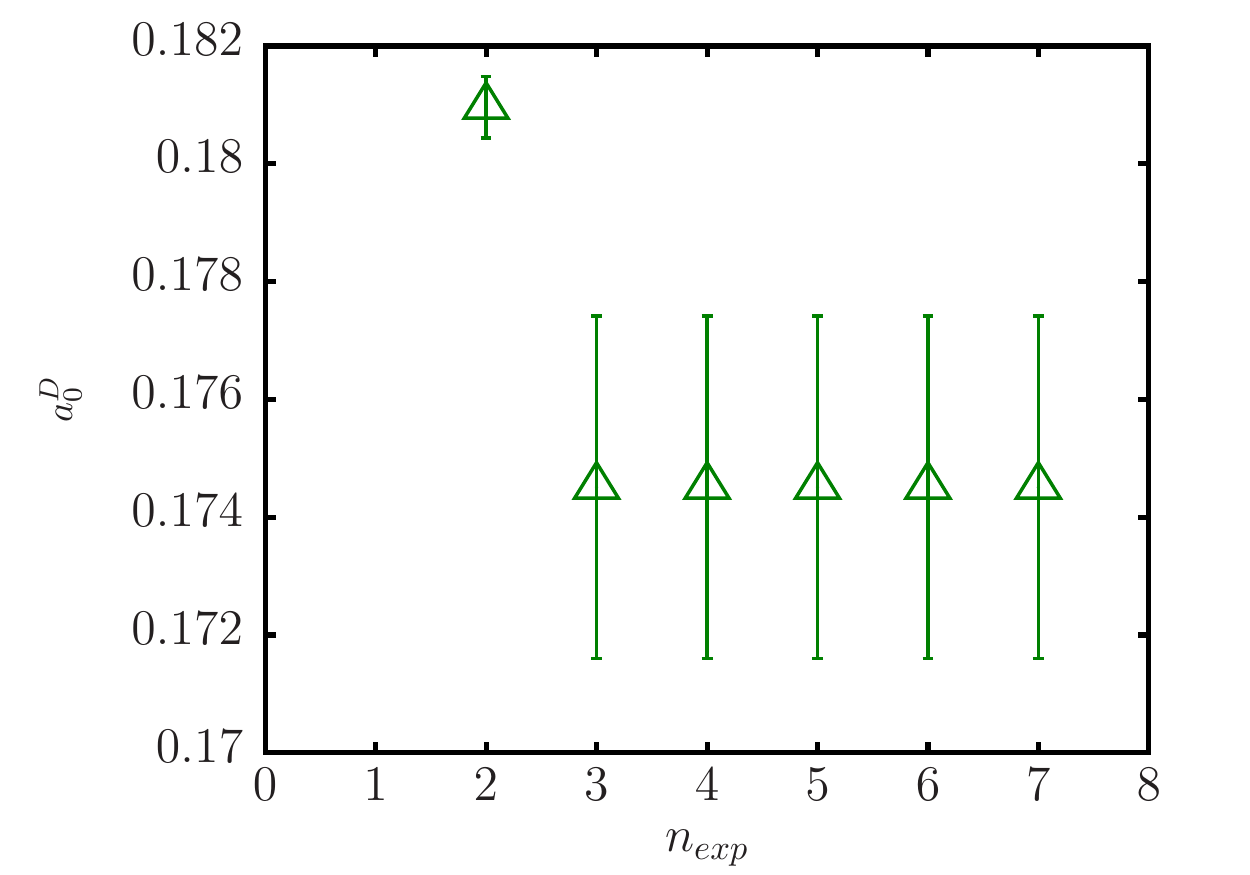}
\end{figure}

While extracting meson ground state properties, we have fitted starting from number of exponentials $n_{\mathrm{exp}}=2$ up to $n_{\mathrm{exp}}=7$ to get a stable fit with a $\chi^2/{\mathrm{dof}} < 1$. For the $D$ and $K$ meson properties, we have achieved stable fit results from the 3rd exponential fits and hence, these results are taken as the final results. The behavior of these results with number of exponentials is shown in Figure \ref{fig:nexpM}.% and \ref{fig:nexpa0}.     
%This huge number ($\sim 200$) of parameters are sometimes difficult to fit using the simultaneous fitting. Therefore I have also tested with sequential fitting where the two-point results are fed as priors to the three-point fits and thus give better stability with much less difficulty. The results extracted from both types of fits are consistent with each other. 

\begin{figure}
\caption{Mass difference of the Goldstone and $\gamma_5\gamma_0$ non-goldstone $D$ mesons.}
\label{fig:massdiff}
\centering
\includegraphics[width=0.5\textwidth]{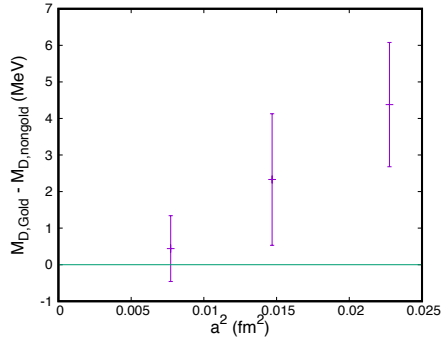}
\end{figure}

We have also tested the taste-effects between the Goldstone and $\gamma_5\gamma_0$ non-Goldstone $D$ mesons arising from the staggered formalism and as expected their mass difference became zero in the continuum, as shown in Figure \ref{fig:massdiff}.

\begin{figure}
\caption{Check for relativistic dispersion relation on all lattice ensembles: the square of the speed of light in free space $c^2$ vs. the kaon momentum $\vec{p}_K$.}
\label{fig:c}
\centering
\includegraphics[width=0.5\textwidth]{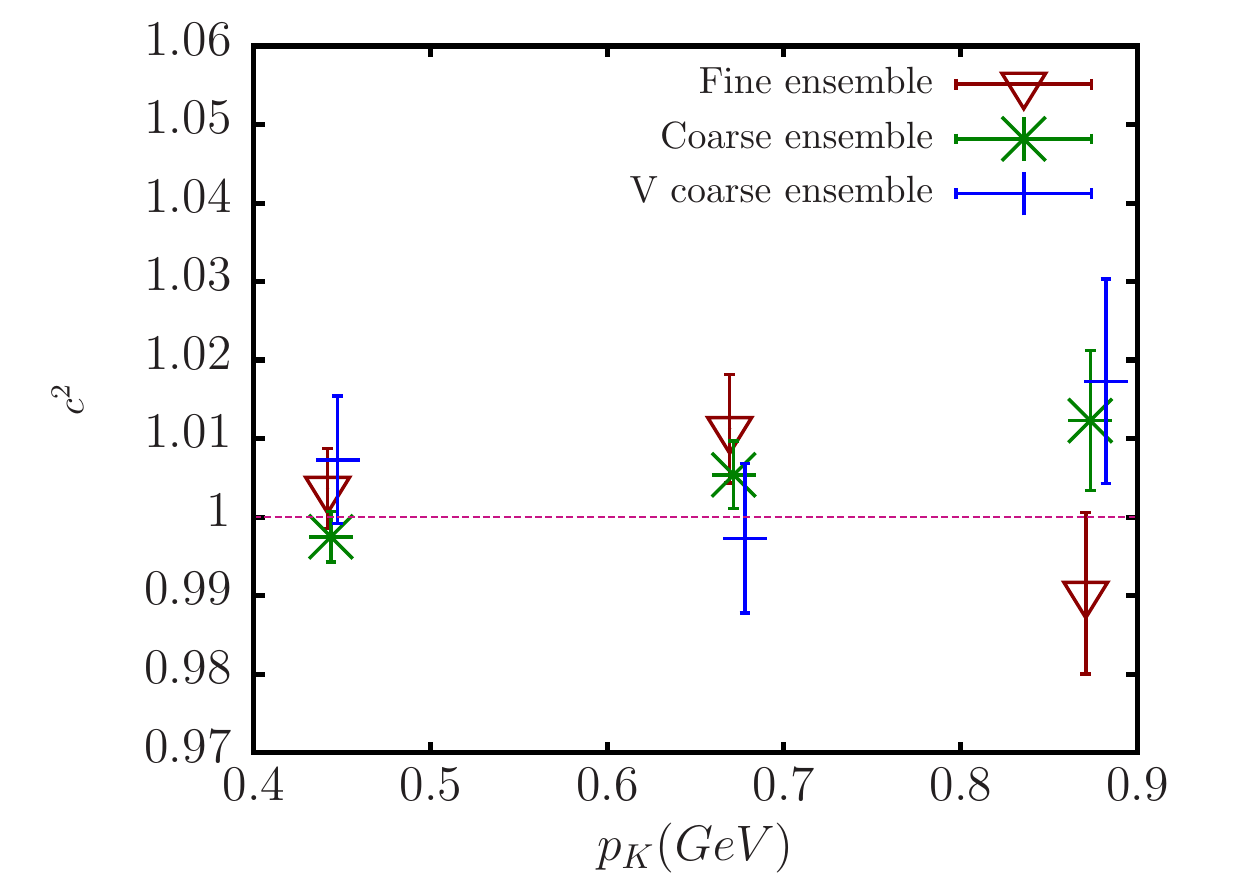}
\end{figure}

We have tested the relativistic dispersion relation as it is only approximate on the lattice. We check the deviation of the square of the velocity of light
\begin{equation}
c^2(\vec{p}) = \frac{E^2_K(\vec{p})-M^2_K}{{\vec{p}}^2}\label{eq:disprel}
\end{equation}
from $1$ for different kinematics and on all ensembles. Generally, on the lattice we expect to get violations $\mathcal{O}(\alpha_S{(pa)}^2)$ for the HISQ formalism. Figure \ref{fig:c} shows that in our calculation we do not see any significant deviation and the relativistic dispersion relation holds within $1-2\%$ statistical deviation, which is within our expectation using the HISQ formalism. However the statistical uncertainties increase in the fitted results for the kaon energies with non-zero momenta.

\begin{figure}
\caption{This plot shows the $q^2$ dependence of the form factors $f_0(q^2)$ and $f_+(q^2)$ with $q^2$ on all the lattice ensembles we have used. These results come from an uncorrelated fit and so is only preliminary at this stage.}
\label{fig:ffvc}
\centering
\includegraphics[width=0.5\textwidth]{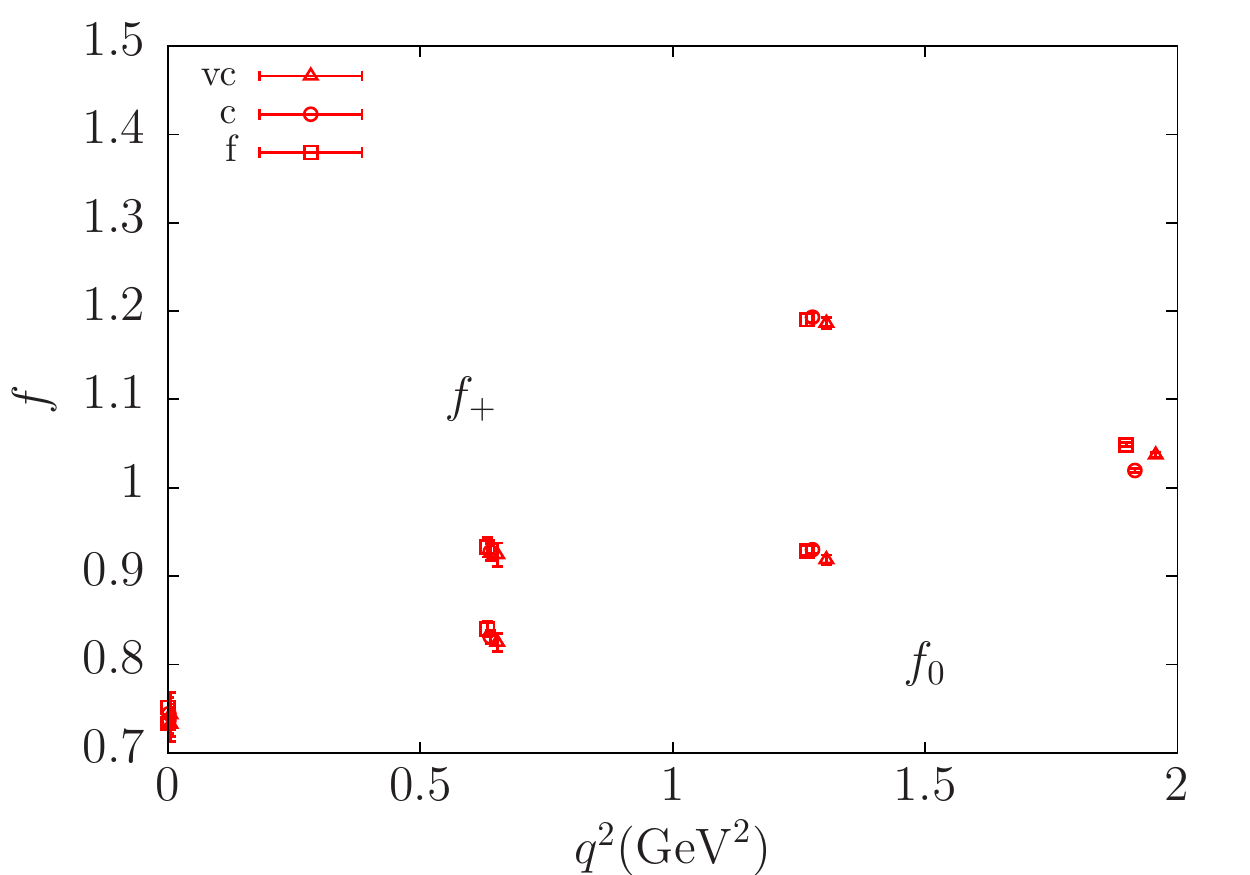}
\end{figure}

The scalar and vector form factors are extracted from the simultaneous fits of all data - including two-point and three-point correlators for all $q^2$ values on each ensemble. Generally the vector current is noisier and hence the vector form factor $f_+(q^2)$. The results for the form factors and their $q^2$ dependence is shown in Figure \ref{fig:ffvc}. These results come from an uncorrelated fit and so is only preliminary at this stage.

%We plan to extrapolate the $q^2$ behavior of the form factors using a modified Z-extrapolation following the references~\cite{Na:2010uf,Koponen:2013tua} which allows for dependence on the lattice spacing and on the light quark mass. And then we can calculate $V_{cs}$ from each experimental $q^2$ bin including bin-to-bin correlation and using equation \label{eq:diffdecayrate} with the experimental decay rates. 
 
%Currently, we are also calculating the form factors on heavier pion mass ensembles to test the light quark mass dependence of the results, and because the results are statistically more precise at heavier light quark masses. 
The results we show here include u/d quarks with physical masses. We plan to extend the study to heavier u/d masses, however, in order to map out the light quark mass dependence. This may also improve our uncertainties somewhat since heavier u/d masses typically give smaller statistical errors. We will then fit our results to a power series expansion in $z$-space (converting from $q^2$ to $z$) following [9]. Taking the coefficients in the $z$-expansion to depend on lattice spacing and light quark mass allows a smooth connection to the continuum physical point where we can compare to experiment, $q^2$-bin by $q^2$-bin, to optimise the final uncertainty on $V_{cs}$. 

%%%%%%%%%%%%%%%%%%%%%%%%%%%%%%%%%%%%%%

\section*{Acknowledgement}

We are grateful to the MILC collaboration for the use of their configurations. Computing was done on the Darwin supercomputer at the University of Cambridge as part of STFC's DiRAC facility. We are grateful to the Darwin support staff for assistance. Funding for this work came from the Gilmour bequest to the University of Glasgow, the National Science Foundation, the Royal Society, the Science and Technology Facilities Council and the Wolfson Foundation. B. C. is supported by the U.S. Department of Energy Office of Science, Office of Nuclear Physics under contract DE-AC05-06OR23177.

\clearpage
\bibliography{lattice2017}

%%%%%%%%%%%%%%%%%%%%%%%%%%%%%%%%%%%%%%%%%%%%%%%%%%%%%%%%%%%%%%%%%%%%%%%%%%%%%
\end{document}